\documentclass[fleqn,twoside]{article}
\usepackage{espcrc2,psfig}

\newcommand{\AmS}{{\protect\the\textfont2
  A\kern-.1667em\lower.5ex\hbox{M}\kern-.125emS}}

\title{Excitation Spectra in a Heavy-Light Meson-Meson System
\thanks{This material is based upon work supported by the
National Science Foundation under Grant No. PHY-0073362.}}

\author{H. R. Fiebig\address{Physics Department, 
                            FIU-University Park, 
                            Miami, Florida 33199, USA} (LHPCollaboration)}
       
\begin{document}

\begin{abstract}
 A system of two static quarks, at
 fixed distances r, and two light quarks is
 studied on an anisotropic lattice. Excitations
 by operators emphasizing quark or gluon degrees
 of freedom are examined. The maximum entropy
 method is applied in the spectral analysis.
 These simulations ultimately aim at learning
 about mechanisms of hadronic interaction.
\end{abstract}

\maketitle

\section{INTRODUCTION}

While QCD has emerged as the foundation of nuclear physics the question of
how the strong hadronic interaction arises from first principles still 
awaits explanation.
Lattice hadron physics holds the most promise for answers. 
Mechanisms of hadronic interaction may
already be studied in systems of hadrons with one heavy, static,
quark each. In this case the relative distance between the hadrons is
well defined, and studies of the interaction mechanism in terms of
intuitive descriptions, like potentials, becomes possible.

In principle, knowledge of the excited states of a two-hadron system
allows the construction of an effective interaction. In practice, the extraction
of excited states from a lattice simulation is a very difficult problem.
The canonical approach is to use plateau fits to effective mass functions.
Bayesian inference \cite{Jar96}, a pillar in many fields of science,
has surprisingly been ignored by the lattice community until only very
recently \cite{Nak00}.
While the aim of this work is to learn about hadronic interaction
mechanisms extraction of the spectral mass density via Bayesian methods
emerges as an interesting subject in its own.

\section{HEAVY-LIGHT MESON-MESON OPERATORS}

\begin{figure}
\psfig{figure=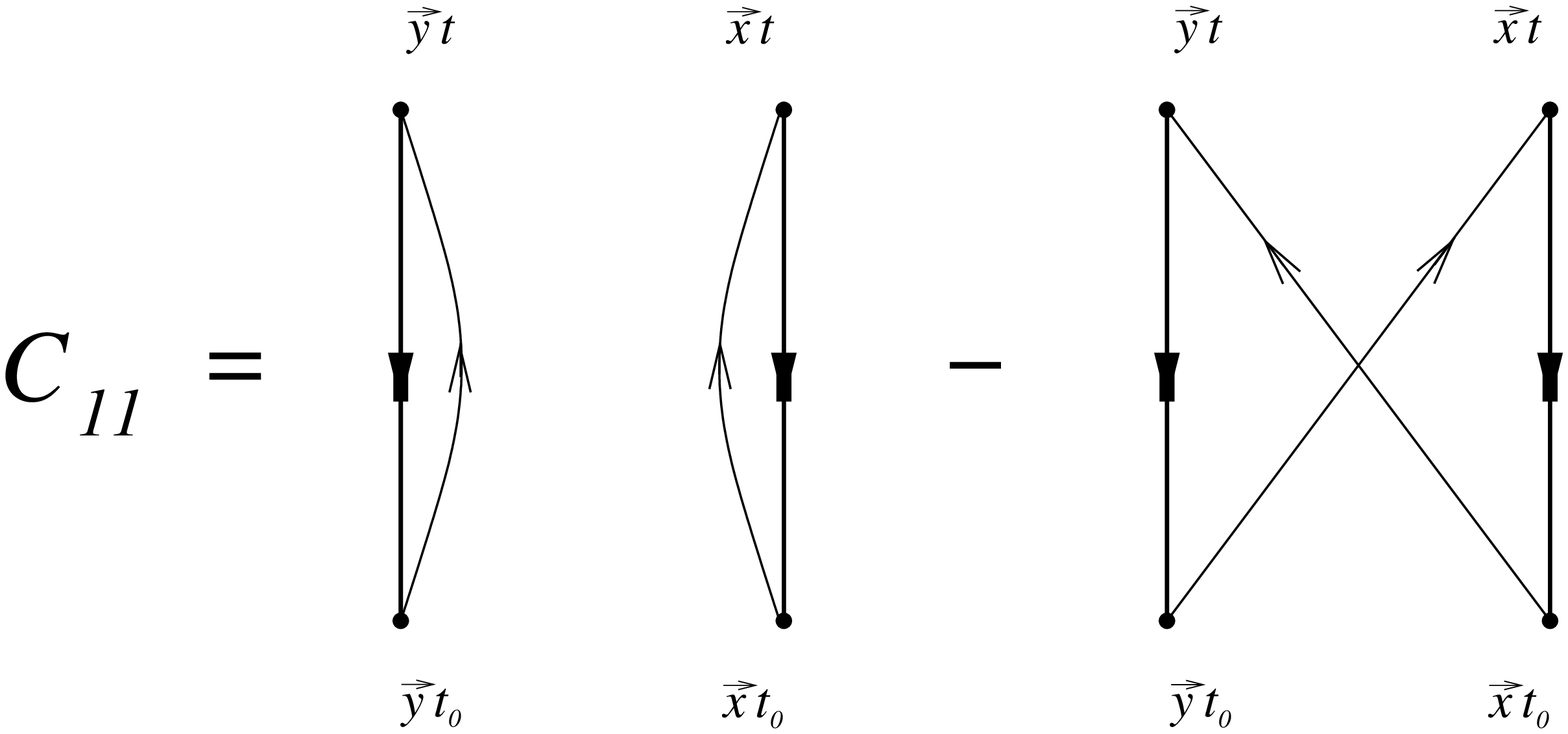,width=54mm,angle=0}\vspace{2ex}
\psfig{figure=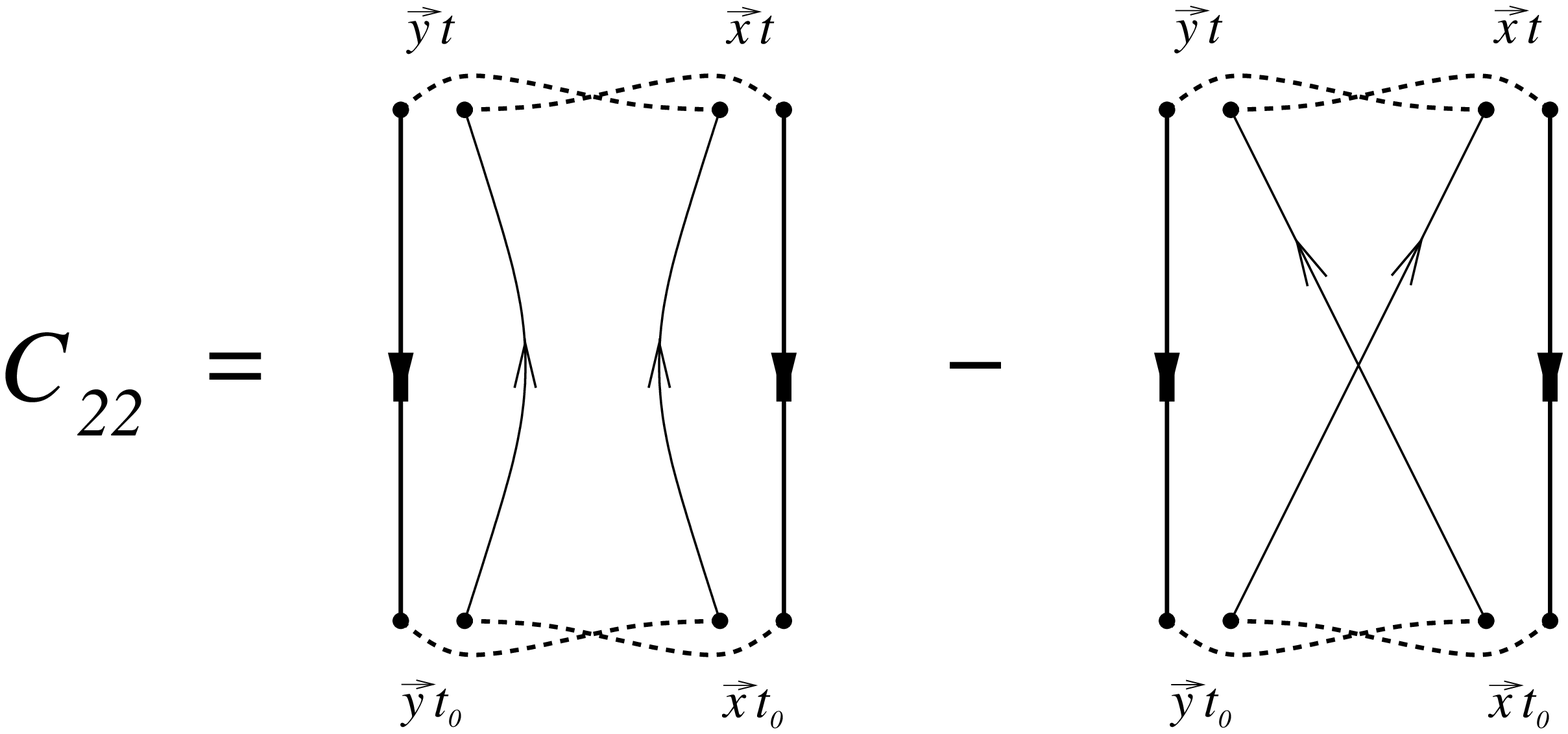,width=54mm,angle=0}
\vspace{-3ex}
\caption{Diagrammatic impressions of the correlators $C_{11}$ and $C_{22}$
representing quark and gluon exchange degrees of freedom, respectively.}
\label{fig2} \end{figure}

Our intention is to test the strengths of different excitation mechanisms,
in a two-meson system, as a function of the relative distance $r$. 
The following operators are used
\begin{eqnarray}
\begin{picture}(0,0)(0,0)
{ \put(50,-32){\oval(38,10)[b]} }
{ \put(121,-32){\oval(38,10)[b]} }
{ \put(87,-108){\oval(112,12)[b]} }
{ \put(85,-106){\oval(33,10)[b]} }
\end{picture}
\lefteqn{\Phi_1(t)=\sum_{\vec{x},\vec{y}}
\delta_{\vec{r},\vec{x}-\vec{y}}} \label{Phi1}\\
& &\overline{Q}_{A}(\vec{x}t) \gamma_5 q_{A}(\vec{x}t)\,
\overline{Q}_{B}(\vec{y}t) \gamma_5 q_{B}(\vec{y}t) \nonumber\\
\rule{0mm}{2mm}\nonumber\\
\lefteqn{\Phi_2(t)=\sum_{\vec{x},\vec{y}}
\delta_{\vec{r},\vec{x}-\vec{y}}} \label{Phi2}\\
&& U_{P; AA^\prime}(\vec{x}t,\vec{y}t)\,
U^\dagger_{P; B^\prime B}(\vec{x}t,\vec{y}t) \nonumber\\
&& \overline{Q}_{ A}(\vec{x}t) \gamma_5 q_{ B}(\vec{x}t)\,
\overline{Q}_{ B^\prime}(\vec{y}t) \gamma_5 q_{ A^\prime}(\vec{y}t)\,.\nonumber
\end{eqnarray}
\rule{-2mm}{6mm}
The operator $\Phi_1$ represents a product of two local pseudo scalar
mesons. The fields $Q$,$q$ denote heavy and light quarks, respectively.
Indices $A,A^\prime,B,B^\prime$ mean color, and $U_P$ are link variable
products along straight paths $P$.
In $\Phi_2$ color singlets, i.e. mesons, spread across a distance $r$.
Excitations due to these nonlocal operators involve gluon exchange degrees of
freedom. In contrast, $\Phi_1$ excitations involve states where the mesons mostly
communicate via $\bar{q}q$ exchange, at large $r$. Diagrammatic representations
of diagonal correlator matrix elements
$C_{ii}(t,t_0)=\langle\hat{\Phi}_i^\dagger(t)\hat{\Phi}_i(t_0)\rangle$
shown in Fig.~\ref{fig2} illustrate these points.
We will loosely refer to $\Phi_1$ and $\Phi_2$ as emphasizing
gluon and quark degrees of freedom (d.o.f.).

\section{TECHNICAL LATTICE ISSUES}

Light-quark propagators are computed from random $Z_2$-source
estimators \cite{Don94} which are non-zero on time slice $t_0$ only.
Heavy-quark propagators
are treated in the static approximation,
i.e. as link products along straight temporal paths.
Gaussian smearing of the quark fields and APE fuzzing of the gauge field
links is applied.

Simulations are done on a $L^3\times T=10^3\times 30$ anisotropic lattice
with a (bare) aspect ratio of $a_s/a_t=3$.
We use the tadpole improved gauge field action of \cite{Mor99} with
$\beta=2.4$. This corresponds to a spatial lattice constant of
$a_s\simeq 0.25{\rm fm}, a_s^{-1}\simeq 800{\rm MeV}$.
The (quenched) anisotropic Wilson fermion action is augmented with
a clover action limited to spatial planes. Only spatial directions are
improved with renormalization factors
$u_s=\langle\, \framebox(5,5)[t]{}\,\rangle^{1/4}$, while $u_t=1$
in the time direction.
The Wilson hopping parameter $\kappa=0.0679$ leads to the mass
ratio $m_\pi/m_\rho \simeq 0.75$.

\section{SPECTRAL DENSITY}

For an operator combination $\Phi_v=v_1\Phi_1+v_2\Phi_2$
the correlation function
\begin{eqnarray}
C_v(t,t_0)&=&\langle\hat{\Phi}^\dagger_v(t)\hat{\Phi}_v(t_0)\rangle =\\
&=&\sum_{n\neq 0}
|\langle n|\hat{\Phi}_v(t_0)|0\rangle|^2
e^{-\omega_n(t-t_0)} 
\end{eqnarray}
has forward ($\omega_n>0$) and backward ($\omega_n<0$) going
contributions on a periodic lattice.
We normalize those differently, defining
\begin{equation}
\exp_T(\omega,t)=\Theta(\omega)e^{-\omega t}
+\Theta(-\omega)e^{+\omega(T-t)}\,,
\end{equation}
the spectral model
\begin{equation}
F(\rho_T|t,t_0)=\int_{-\infty}^{+\infty}d\omega\,
{\rho_T(\omega)}\exp_T(\omega,t-t_0)
\label{Fc}\end{equation}
is then expected to fit the lattice correlators.
Specifically, the requirement $F(\rho_T|t,t_0)=C_v(t,t_0)$ leads to
a sum of discrete $\delta$-peaks for the spectral density 
\begin{eqnarray}
\rho_T(\omega)&=&
\sum_{n\neq 0}\delta(\omega-\omega_n)\,
|\langle n|\hat{\Phi}_v(t_0)|0\rangle|^2\times \\
&&[\Theta(\omega_n)+\Theta(-\omega_n)e^{-\omega_n T}]\,. \nonumber
\end{eqnarray}
The peak-like signature should survive in the discretized form of (\ref{Fc}) 
\begin{equation}
F(\rho|t,t_0)\simeq\sum_{k=K_-}^{K_+}\rho_k\,\exp_T(\omega_k,t-t_0)\,,
\end{equation}
where $\Delta\omega$ is a suitable interval,
$\omega_k=\Delta\omega k$,
and $\rho_k=\Delta\omega\rho_T(\omega_k)$.

From the Bayesian point of view the spectral weights $\rho_k, k=K_-\ldots K_+$,
as well as the
values $C_v(t,t_0), t=0\ldots T-1$, of the
correlation function are stochastic variables, $\rho$ and $C$, subject
to certain probability distribution functions.
The conditional probability ${\cal P}[C\leftarrow \rho]$ of
$C$, given $\rho$, is known as the likelihood function.
For a large number of measurements we have
${\cal P}[C\leftarrow \rho]\propto e^{-\chi^2/2}$,
see \cite{Jar96}, where
\begin{eqnarray}
\lefteqn{ \chi^2 = \sum_{t_1,t_2}
\left[ C_v(t_1,t_0)-F(\rho_T|t_1,t_0)\right]\times } \\
&&\Gamma^{-1}_v(t_1,t_2)\,
\left[ C_v(t_2,t_0)-F(\rho_T|t_2,t_0)\right] \nonumber
\end{eqnarray}
is the usual $\chi^2$-distance between the data and the model,
and $\Gamma^{-1}_v(t_1,t_2)$ denotes the inverse covariance
matrix \cite{Bra76}. It accounts for the statistical dependence
of correlator data between time slices.
Assuming minimal information (in the sense of \cite{Sha49}) about
$\rho$ the Bayesian prior probability
is $P[\rho]\propto e^{\alpha S}$ where $\alpha$ is a parameter and
\begin{equation}
S[\rho] = \sum_k \left( \rho_k-m_k-\rho_k\ln\frac{\rho_k}{m_k}\right)
\end{equation}
is the entropy relative to $m$, the latter being called the default model \cite{Jar96}.
By virtue of Bayes' theorem \cite{Box73} the so-called posterior
probability is
\begin{equation}
{\cal P}[\rho\leftarrow C] = {\cal P}[C\leftarrow\rho]P[\rho]/P[C]
\propto e^{-W[\rho]}\,,
\end{equation}
for a given $C$, where $W$ is defined as
\begin{equation}
W[\rho]=\chi^2/2-\alpha S\,.
\label{Wrho}\end{equation}
The most likely $\rho$ is obtained by minimizing $W[\rho]$. The method
of choice \cite{Nak00}, so far, has been singular value decomposition (SVD).
Indeed, the functional $W[\rho]$ has a unique minimum \cite{Jar96}.
Stochastic methods, however, seem more in tune with the Bayesian perspective.
In this spirit we have employed simulated annealing to obtain
the spectral density.
The corresponding partition function is
\begin{equation}
Z_W=\int[d\rho]\,e^{-\beta_W W[\rho]}\,.
\end{equation}
A Metropolis algorithm was used with local updates $\rho_k\rightarrow x\rho_k$
where $x$ is a random deviate with distribution $p_2(x)=xe^{-x}$.
For the annealing schedule a power law
$\beta_W(n)=(\beta_1-\beta_0)\left(n/N\right)^\gamma+\beta_0$
turned out to be suitable.

\section{RESULTS}

\begin{figure}[htb]
\psfig{figure=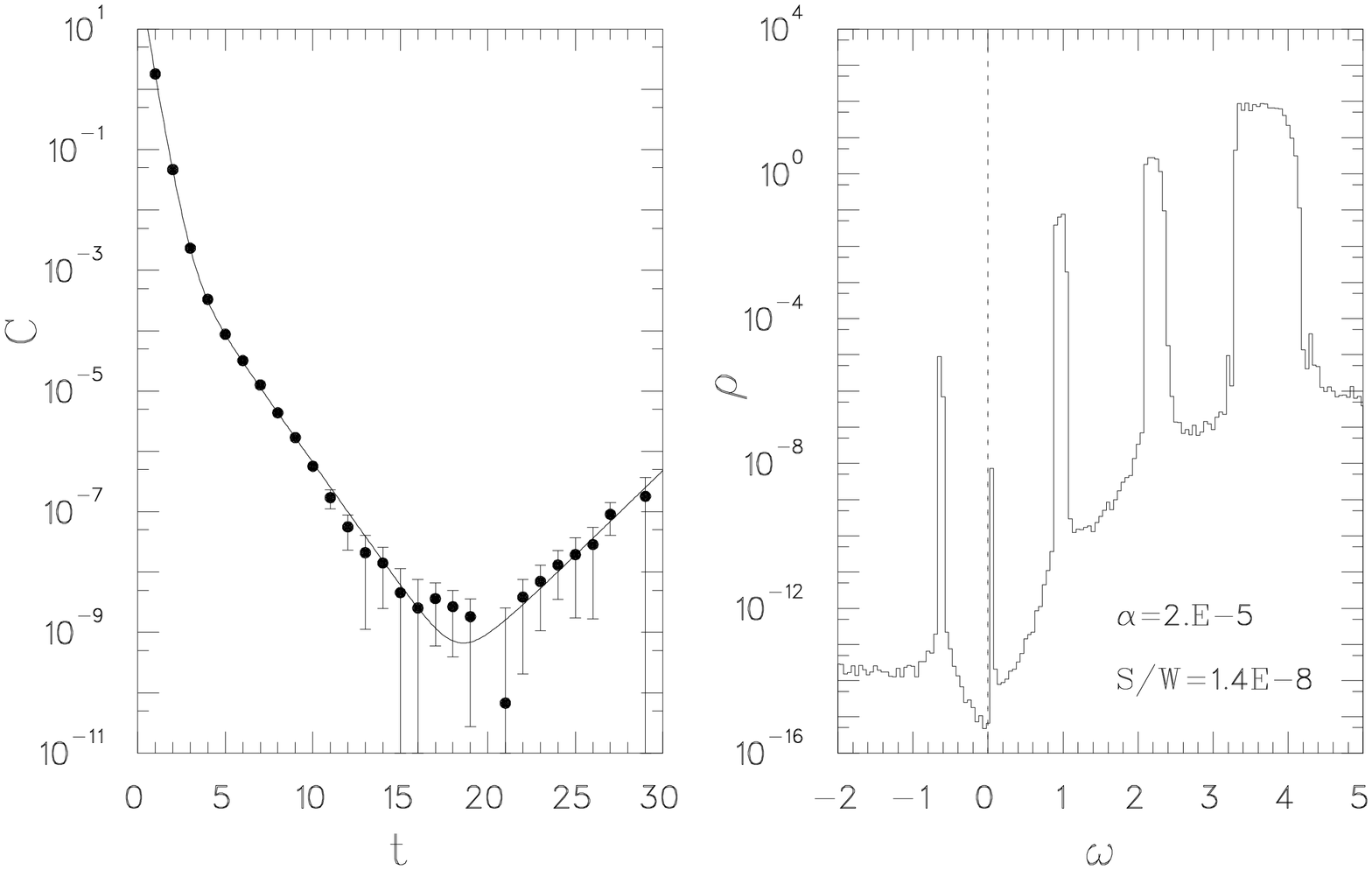,width=74mm,angle=90}\vspace{2ex}
\psfig{figure=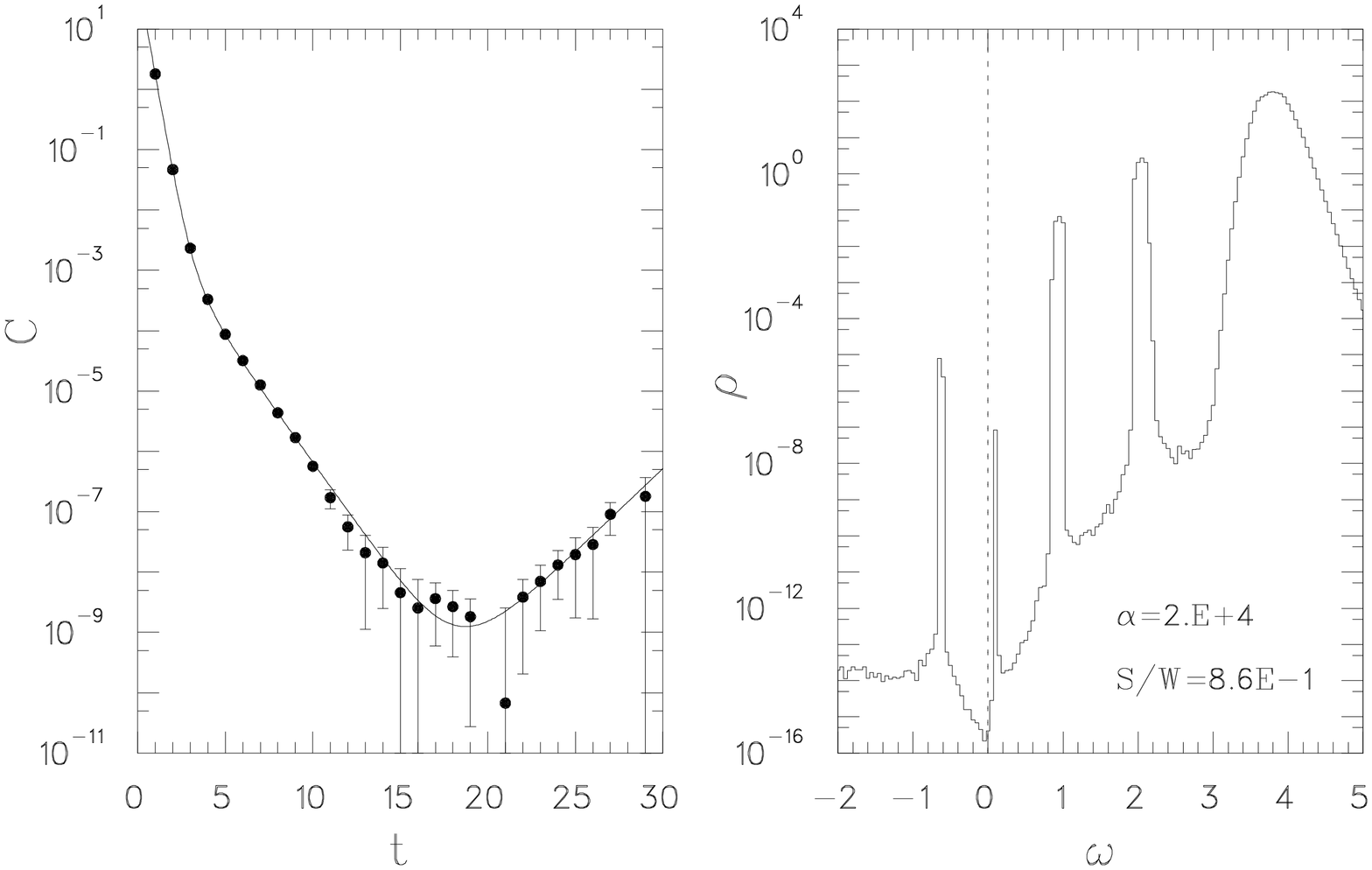,width=74mm,angle=90}
\vspace{-3ex}
\caption{Entropy weight dependence.
Correlators and spectral densities for an artificial
data set, at relative distance $r=1$ and constant default model $m=10^{-12}$,
for $\alpha=2\times 10^{-5}$ and $2\times 10^{+4}$, respectively.}
\label{fig3} \end{figure}

We have explored the $\alpha$-dependence of $\rho$ using an artificial (mock)
data set compiled from eigenvalues of the $2\times 2$ correlation matrix 
$C_{ij}(t,t_0)$.
Cooling averages were taken during the last 200, of $N=2200$,
cooling steps during which the final $\beta_W$ was kept constant.
For tuning purposes it is useful to monitor the entropy load
\begin{equation}
S/W:=\frac{\langle -\alpha S\rangle_{\beta_W\rightarrow\infty}}
{\langle W\rangle_{\beta_W\rightarrow\infty}}\,.
\label{SW}\end{equation}
At constant default model $m=10^{-12}$ the entropy weight
$\alpha$ was varied over 15 orders of magnitude.
Two examples are shown in Fig.~\ref{fig3}.
Remarkably the gross structure of the spectral density remains
stable across 8 orders of magnitude of $\alpha$. The second pair
of frames in Fig.~\ref{fig3} marks a kind of critical point $\alpha_S$,
beyond which the annealing action $W$ becomes entropy dominated,
the onset of ignoring the evidence ${\cal P}[C]$. This results in a
smoothing of the contours of the large-$\omega$ peak, see Fig.~\ref{fig3}. 
Even so, the fit to the correlation function is not greatly affected.
We found that a good criterion for choosing the entropy weight
parameter $\alpha$ is to tune $S/W$ to about $\leq 10^{-2}$.
As is evident from Tab.~\ref{tab1}, near this value the linear relationship
between $S/W$ and $\alpha$ begins to break down.
\begin{table}[htb]
\caption{The entropy load $S/W$ for some entropy weight parameters $\alpha$.
Note the linear relationship in the region $\alpha\leq 2\times 10^{+1}$.} 
\vspace{2ex}
\begin{tabular}{@{}cc}\hline
$\alpha$ & $S/W$ \\ \hline
$2\times 10^{-5}$ & $1.4\times 10^{-8}$ \\
$2\times 10^{-2}$ & $1.4\times 10^{-5}$ \\
$2\times 10^{+1}$ & $1.4\times 10^{-2}$ \\
$2\times 10^{+4}$ & $8.6\times 10^{-1}$ \\ \hline 
\end{tabular}
\label{tab1}\end{table}

\begin{figure}[htb]
\psfig{figure=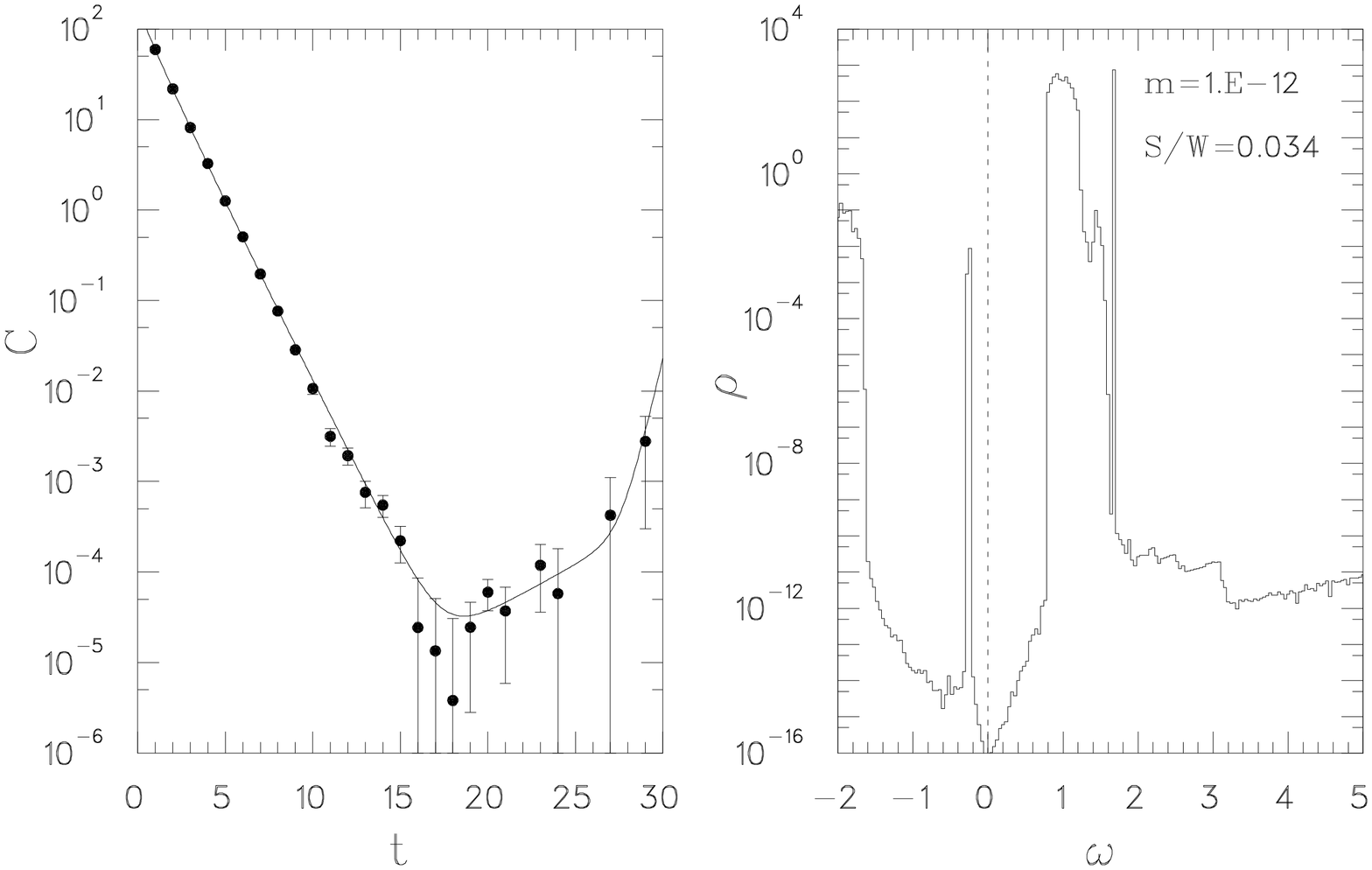,width=74mm,angle=90}\vspace{2ex}
\psfig{figure=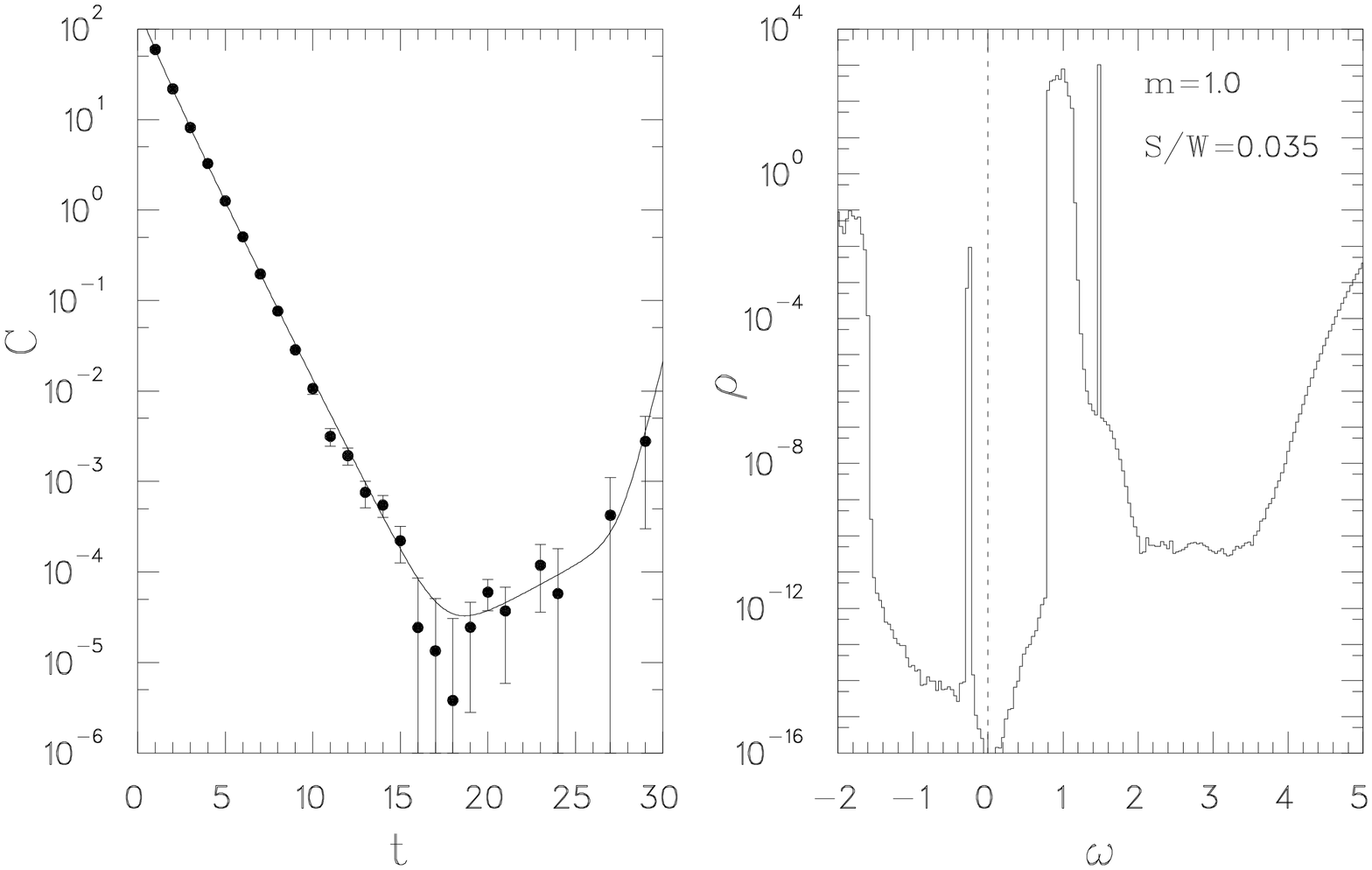,width=74mm,angle=90}
\vspace{-3ex}
\caption{Default model dependence.
Correlators of the two-meson operator $\Phi_1$
at relative distance $r=2$. The spectral densities are obtained with
constant default models of $m=10^{-12}$ and $m=1.0$, respectively,
at constant entropy load $S/W$.}
\label{fig4} \end{figure}

To explore the default model dependence the correlation function $C_{11}$
of the local meson-meson operator (\ref{Phi1}) was analyzed for a series
of constant models $m$. The results for two values, $m=10^{-12}$ and $m=1.0$,
are shown in Fig.~\ref{fig4}.
Different entropy weight
parameters, $\alpha=100$ and $1000$, were chosen such that the entropy load
$S/W$ is about the same.
Increasing $m$ by twelve orders of magnitude changes $\rho$ at the
high-$\omega$ end. In that region spectral peaks are apparently not supported
by the data. Thus the spectral density tends towards the default model.
Again, absence of information leads to a smoother curve.
Note that the change in $\rho$ for $\omega>3$ is numerically
inconsequential due to the presence of large peaks,
also, the gross structure of the spectral function is stable against changing $m$. 
On the other hand,
the micro structure of like peaks in Fig.~\ref{fig4} differs noticeably.

\begin{figure}[htb]
\psfig{figure=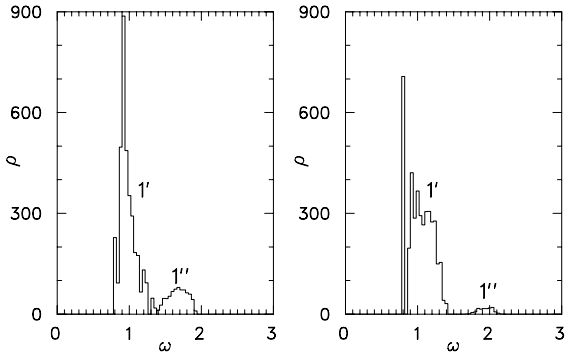,width=74mm,angle=90}
\vspace{-3ex}
\caption{Spectral densities from a cold and a random start,
for $\Phi_1$ at $r=2$, also see Tab.~\protect\ref{tab2}.}
\label{fig5} \end{figure}

It is well known that the functional (\ref{Wrho}) has a unique
absolute minimum \cite{Jar96}. Any start configuration used in the annealing
process will result in a final configuration $\rho$ that is merely close to the
absolute minimum, since $\beta_W<\infty$.
Then, which features of $\rho$ are independent of the start configuration?
Figure~\ref{fig5} shows two spectral functions obtained from a cold
start at $m=10^{-12}$ and a random start. 
There is a main and a secondary peak ($1^{\prime}$ and $1^{\prime\prime}$).
Repeated computation of $\rho$
with varying random starts and annealing schedules shows that the
two-peak feature is stable, thus being a property of the data.
We have looked at peak volumes and energies
\begin{eqnarray}
Z_n&=&\int_{\delta_n}d\omega\,\rho_T(\omega)\;=\;
|\langle n|\hat{\Phi}_v(t_0)|0\rangle|^2 \label{Zn} \\
E_n&=&Z_n^{-1}\int_{\delta_n}d\omega\,\rho_T(\omega)\,\omega \label{En}
\end{eqnarray}
where $\delta_n=\{\omega\,|\,\omega\in \mbox{peak \#$n$}\}$.
In Tab.~\ref{tab2} there are listed the peak volumes obtained from a
sample of 4 starts (1 cold, 3 random), and the averages (ave) and standard deviations (sig).
Assuming that the spectral density comprises two distinct states,
$1^{\prime}$ and $1^{\prime\prime}$, the variances of
$Z_{1^\prime}$ and $Z_{1^{\prime\prime}}$ are $\approx 10$.
Their combined volume $Z_{1^\prime\cup\,1^{\prime\prime}}$ however has a variance
less by one order of magnitude.
Peak splits, and also spikes, can be caused by imperfect lattice data.
In this light, interpreting Fig.~\ref{fig5} in terms of only one physical state,
$1=1^\prime\cup\,1^{\prime\prime}$, as opposed to two, $1=1^\prime$ and $2=1^{\prime\prime}$,
is a possibility. More analysis work will be needed to decide this matter.
\begin{table}[htb]
\caption{A possible peak split. With 4 different annealing starts
the peak volume (\protect\ref{Zn}) fluctuates between peaks ${1^\prime}$ and $1^{\prime\prime}$.}
\vspace{2ex}
\begin{tabular}{@{}cccc}\hline
Start & $Z_{1^\prime}$ & $Z_{1^\prime\cup\,1^{\prime\prime}}$ & $Z_{1^{\prime\prime}}$ \\ \hline
  1 &  142.0 & 170.9 & 28.9 \\
  2 &  162.9 & 167.9 & 5.0 \\
  3 &  155.4 & 170.2 & 14.8 \\
  4 &  141.1 & 171.0 & 29.9 \\ \hline
  ave & 150.3 &  170.0 & 19.7 \\
  sig & 9.2 &  1.3 & 10.4 \\ \hline
\end{tabular}
\label{tab2}\end{table}

\begin{figure}
\psfig{figure=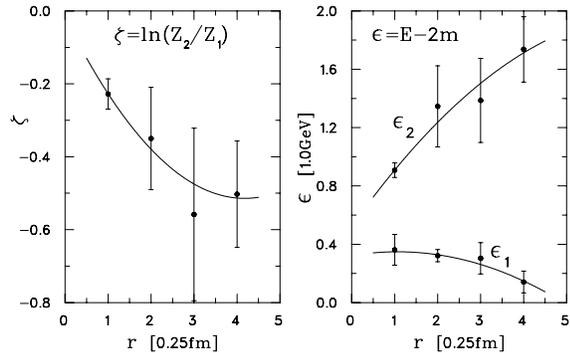,width=74mm,angle=90}
\vspace{-3ex}
\caption{Normalization independent excitation strength ratio $\zeta$ of gluon versus
quark exchange degrees of freedom as a function of the relative
meson-meson distance $r$ (left).
Also shown are the corresponding energy shifts $\epsilon_{1,2}$
compared to the mass $2m$ of two noninteracting mesons (right).
The error bars reflect annealing start standard deviations.}
\label{fig6} \end{figure}

Amoung the features of $\rho$ that are computable via spectral
analysis are low-$\omega$ moment integrated quantities, like the peak
volume $Z_n$ and energy $E_n$, for some state $n$.
These are interesting observables because they may reveal aspects of the
physics of hadronic interaction. For example, the relative strength
of excitations due to operators $\Phi_2$ versus $\Phi_1$ is measured by
$\zeta=\ln(Z_2/Z_1)$, in a normalization independent way.
We have plotted $\zeta$ in Fig.~\ref{fig6}   
as a function of the relative meson-meson distance $r$.
As $r$ decreases $\Phi_2$-type excitations gain strength over
$\Phi_1$-type excitations.
In view of (\ref{Phi1}), (\ref{Phi2}), and Fig.~\ref{fig2} we may say that gluon exchange
degrees of freedom move to dominate quark exchange d.o.f. at small $r$.
We consider Fig.~\ref{fig6} a preliminary result, a more detailed analysis is in progress.
Possibly, in the chiral limit, level crossing of $\epsilon_2$ and $\epsilon_1$
in Fig.~\ref{fig6} may occur \cite{Fie00a}.   

\section{CONCLUSION}

Studies of hadronic interaction require extracting excited states from lattice
simulations. Bayesian inference proves a superior tool compared to standard
plateau methods. Using an entropic Baysian prior integrated quantities, like
the excitation strengths and energies of certain states, can be reliably computed.
The excitations of a heavy-light meson-meson give insight into
aspects of strong interaction physics.
At the current lattice parameters, quark degrees of freedom
still dominate the physics of hadronic interaction.

\end{document}